\documentstyle[12pt,aaspp4]{article} 

\newcommand{\etal} {et~al.\ } 
\newcommand{\HI}{H{\sc i}} 
\newcommand{\HII}{H{\sc ii}}
\def\deg{\ifmmode^{\circ}\else$^{\circ}$\fi} % overwrites \deg in LaTeX
\def\min{\ifmmode^{\prime}\;\else$^{\prime}$\fi}
\def\sec{\ifmmode^{\prime\prime}\;\else$^{\prime\prime}\;$\fi}
\newcommand{\kms}{${\rm km~s^{-1}}$}

\lefthead{McClure-Griffiths \etal} 
\righthead{Large HI Shells near l=279} 

\begin{document} 

\title{Two Large \HI\ Shells in the Outer Galaxy near {\boldmath $l=279$\deg}}
\author{N. M. McClure-Griffiths\altaffilmark{1} \& John M. Dickey}
\affil{Department of Astronomy, University of Minnesota, 116 Church Street
SE, Minneapolis, MN 55455} 
\author{B. M. Gaensler\altaffilmark{2}}
\affil{Center for Space Research, Massachusetts Institute of Technology, 70 Vassar Street, Cambridge, MA 02139}
\author{A. J. Green}
\affil{Astrophysics Department, School of Physics, Sydney University, NSW 2006, Australia}
\author{R. F. Haynes \& M. H. Wieringa} 
\affil{Australia Telescope National Facility, CSIRO, P.O. Box 76, Epping, NSW 2121, Australia}
\altaffiltext{1}{naomi@astro.umn.edu}
\altaffiltext{2}{Hubble Fellow}

\vspace{0.25 in}
%{\it This manuscript has been accepted for publication in the June 2000 issue of the
%{\rm Astronomical Journal.}  No reference should be made to its contents
%prior to publication without permission of the authors.}

\vspace{0.25 in} 
\authoraddr{Address correspondence regarding this manuscript to: 
		N. M. McClure-Griffiths 
		Dept. of Astronomy University of
		Minnesota 116 Church St. S.E.  
		Minneapolis, MN 55455}
%--------------------------------------------
\begin{abstract}
%-------------------------------------------
As part of a survey of \HI\ $\lambda$21-cm emission in the Southern
Milky Way, we have detected two large shells in the interstellar neutral
hydrogen near $l=279^{\circ}$.  The center velocities are +36 and +56 km s$^{-1}$,
which puts the shells at kinematic distances of 7 and 10 kpc.  The larger
shell is about 610 pc in diameter and very empty, with density contrast of
at least 15 between the middle and the shell walls.  It has expansion
velocity of about 20 km s$^{-1}$ and swept up mass of several million solar
masses.  The energy indicated by the expansion may be as high as $2.4 \times
10^{53}$ ergs.  We estimate its age to be 15 to 20 million years.  The
smaller shell has diameter of about 400 pc, expansion velocity about 10 km
s$^{-1}$ and swept up mass of about $10^6$ solar masses.

Morphologically both regions appear to be shells, with high density regions
mostly surrounding the voids, although the first appears to have channels of
low density which connect with the halo above and below the \HI\ layer.
They lie on the edge of the Carina arm, which suggests that they may be
expanding horizontally into the interarm region as well as vertically out of
the disk.  If this interpretation is correct, this is the first detection of
an \HI\ chimney which has blown out of both sides of the disk.
\end{abstract}


%----------------------------------------------------------
\section{Introduction}
\label{sec:intro}
%---------------------------------------------------------
Studies of external galaxies, in particular, recent studies of the Large and
Small Magellanic Clouds, indicate large populations of shells and
supershells, which dominate the structure of the interstellar medium (ISM)
(Staveley-Smith \etal 1997; Kim \etal 1998).  By injecting large quantities
of energy into the ISM, these shells reshape galaxies on size scales of tens
to hundreds of parsecs and trigger new star formation.  In our own Galaxy,
surveys have found many small shells and ``worms'', (eg. Heiles 1979, 1984;
Koo, Heiles, \& Reach 1992 (KHR)) but the number of large supershells and
chimneys which have been identified is still relatively small (Normandeau,
Taylor, \& Dewdney 1996; Heiles 1998). These exceptionally large structures
are most often identified as dramatic voids in the Galactic neutral
hydrogen, observed with the \HI\ line at 1420 MHz.  Unfortunately, in the
inner Galaxy, where they are most likely to occur, they prove difficult to
detect due to distance ambiguities.  As a result, our knowledge of how
dramatically the Galaxy has been shaped by shells is limited.

These \HI\ voids range in size from tens of parsecs to kiloparsecs, and are
found with a variety of morphologies from nearly spherical to chimney-like.
The dominant paradigm suggests that these structures are caused by the
combined pressures of stellar winds and sequential supernovae (SNe) in OB
associations (Heiles 1984).  It has also been suggested, however, that the
largest of these structures, with energies in excess of $10^{53}$ ergs, may
be caused by impacts of high velocity clouds (HVCs) with the Galactic disk
(cf. Heiles 1984 for a discussion of both formation methods), or more
recently, that they are the remnants of hypernovae and/or gamma ray bursts
(Loeb \& Perna 1998).  Younger supershells are often associated with some
ionized emission in the shell interior in the form of hot x-ray emitting
gas, or an H$\alpha$ emitting inter-rim (Points \etal 1999). For the oldest
supershells it is likely that the hot x-ray emitting medium has diffused and
the massive stars of the OB association have expired, leaving only a
evacuated region in the Galactic \HI.  For the largest, and therefore the
oldest, of these shells, expansion can exceed the scale height of the \HI\
layer of the Galaxy.  Such expansion will elongate along the axis
perpendicular to the Galactic plane, as predicted by theories of expansion
into a stratified medium (eg.  Kompaneets 1960).  In this case we expect to
see chimneys where the polar regions of the shell become Rayleigh-Taylor
unstable and break through into the Galactic halo, providing a source of
ionized hydrogen and thermal support for the halo.

The Galactic Plane near $l=280\deg$, $v=+35$ \kms\ (LSR)\footnote{All
velocities are quoted with respect to the Local Standard of Rest (LSR).}  is
a dynamic place, with very dramatic brightness temperature fluctuations over
relatively small scales, and the edge of the Carina arm.  Positive
velocities in this direction are beyond the solar circle, corresponding to a
unique distance.  As a result, it is somewhat easier to unravel the Galactic
structure in this region than it is in the inner Galaxy.  The $l=280\deg$
line-of-sight is tangent to the Carina spiral arm.  The region between
$l=275\deg$ and $l=280\deg$ and $v=+25$ \kms\ and $v=+50$ \kms\ is in
between spiral arms, with the Carina arm towards greater longitudes and the
Perseus arm towards lesser longitudes or higher velocities.  In the \HI\ and
CO longitude-velocity ({\em l-v}) diagram the Carina arms forms a loop with
the apex at $l=280\deg$, $v=0$ \kms.  Towards greater longitudes the Carina
arm is seen to extend along $v\sim 35$ \kms\ to $l=330\deg$ and beyond
(Grabelsky \etal 1987).  At lesser longitudes the Vela supernova remnant
dominates radio continuum and X-ray emission towards $l=265\deg$.

In this paper we report on the discovery of two large Galactic \HI\ shells
near the Carina tangent.  One, GSH 277+0+36, is centered at a galactic
longitude of $l=277$\deg, latitude of $b=0$\deg\, and velocity $v=+36$ \kms\
with an angular diameter of 5\deg.5.  The second, smaller shell, GSH
279+0+59 is centered on $l=280\deg$, $b=0\deg.1$, $v=+59$ \kms\ with an
angular diameter of 2\deg.7.  We will explore the possibility that the
shells are interarm voids as previously suggested (Grabelsky \etal\ 1987),
and present arguments in favor of a shell interpretation.  In
section~\ref{sec:obs} we describe the observations and analysis.  In
sections~\ref{subsec:sh1-morph} and \ref{subsec:sh1-props} we discuss the
morphology and physical properties of GSH 277+0+36.  In
section~\ref{subsec:sh2} we discuss the morphology and properties of GSH
279+0+59.  In section~\ref{subsec:otherwaves} we compare the \HI\ emission
with other wavebands, including far-infrared, 2.4 GHz continuum, and CO.
Finally, in section~\ref{sec:disc} we discuss possible formation methods.

%----------------------------------------------------------
\section{Observations and Analysis}
\label{sec:obs}
%----------------------------------------------------------
The observations were made as part of the Southern Galactic Plane Survey
(SGPS), a large project to map the $\lambda$21-cm continuum and \HI\
spectral line in the fourth quadrant of the Galactic Plane with high angular
and velocity resolution (Dickey \etal 1999; McClure-Griffiths \etal 2000).
The SGPS makes use of high spatial resolution data from the Australia
Telescope Compact Array (ATCA) near Narrabri, Australia, and zero spacing
information from the Parkes 64m radio telescope near Parkes,
NSW.\footnote{The Parkes telescope and the ATCA are part of the Australia
Telescope, which is funded by the Commonwealth of Australia for operation as
a National Facility managed by the Commonwealth Scientific and Industrial
Research Organisation.}  The final project will provide a complete \HI\ and
continuum data set of $253\deg \leq l \leq 358$\deg\ and $-1\deg.0 \leq b
\leq 1\deg.0$ at an angular resolution of 1\min, and with velocity
resolution of $\Delta v = 0.82~{\rm km~s^{-1}}$.  In addition, we have
extended the single dish coverage to $b=\pm 10\deg$ in order to study large
scale structures which protrude from the Galactic Plane.

The observations on 1998 December 15 and 16 covered the Galactic longitude
253\deg\ to 358\deg, with Galactic latitude coverage of $\pm 1\deg.5$.
Between 1999 June 18 and 21 1999 we extended the coverage to $-7\deg.5 \leq
b \leq +4\deg.5$.  We finished the extensions to $b=\pm 10$\deg\ during
observations spanning 1999 September 18-27.

The data presented here were obtained during three observing sessions using
the multibeam receiver package on the Parkes telescope.  The multibeam
system is a thirteen beam focal plane array at $\lambda$21-cm.  It is
comprised of thirteen independent feeds, each with dual, cryogenically
cooled, orthogonal, linear polarization receivers (Staveley-Smith \etal
1996).  The beams are arranged on a hexagonal grid with a 29\min.1
separation between adjacent feeds.  The array was designed to optimize dish
illumination, as a result the array undersamples the focal plane (at
$\lambda 21$-cm the beam FWHP is 14\min.4.  In addition, the feeds have low
system temperatures ($T_{\rm sys} \sim 20$ K) which, in these data, resulted
in a mean rms noise of $\sim 0.3$ K in each channel.

The multibeam correlator is capable of operating in several modes.  There is
a wide band mode as used by the HIPASS and ZOA surveys (Staveley-Smith \etal
1996), as well as a new narrow band mode which operates with 2048 channels
spread across an 8 MHz band (Haynes \etal 1998).  For this survey we operate
in the latter mode in order to match the channel width ($\Delta v = 0.8$
\kms) of the ATCA data.  Due to computing limitations in the correlator, the
narrow band mode is only operated on the inner seven beams of the multibeam
system.

Our observing strategy at Parkes was to use the multibeam receiver to map
``on-the-fly''.  In this technique, the telescope was driven at a rate of
about $1.4~{\rm deg~min^{-1}}$ writing samples every 5 s.  In order to
maximize the sky coverage, and reduce redundant samples, the receiver
platform was continuously rotated at an angle of $19\deg.1$ to the scan
direction as the telescope was scanned through three degrees of Galactic
latitude at a constant longitude.  This resulted in parallel tracks with an
angular separation of 9\min.5.  Because the 9\min.5 spacing is worse than
Nyquist sampling, interleaved scans are necessary.  The resultant map has
parallel tracks spaced 4\min.7 apart.  In the scan direction, samples are
7\min.2 apart.  In order to reduce the effects of gain variations between
feeds, care was taken to ensure that independent feeds were responsible for
adjacent tracks.

In order to reduce the effects of system temperature vs.\ elevation
variations we observed so as to maintain a nearly constant zenith angle ($ZA
\sim 30\deg$).  We were unable to scan at a zenith angle of less than the
$ZA\sim 25\deg$ without overtaxing the azimuth drives on the telescope. The
IAU standard regions S6 and S9, were observed at similar zenith angles as
the rest of the observations for the day.

Frequency switching was carried out during the observations in order to
allow for rigorous off-line calibration.  The spectra were centered on 1419
MHz and 1422.125 MHz, with 10 s integration times.  Each sample was divided
by the previous frequency switched cycle to remove the front-end gain vs.\
frequency shape.  The bandpass shape was then fitted with a series of
Fourier components and the data were divided by the determined shape.
Absolute brightness temperature calibration of the \HI line data was
performed using the standard regions S6 and S9 with standard line
temperatures given in Williams (1973).  The integral over part of the line
was used to determine a calibration scale factor, $C$, for each polarization
on each beam.  The brightness temperature was therefore calculated according
to:
\begin{equation}
 T_{b} = C \times T_{bas} ~\frac{T_{obs}(\nu)}{T_{ref}(\nu)} - T_{sys},
\label{eq:brighttemp}
\end{equation}
where $T_{ref}(\nu)$ is a smoothed version of the frequency switched,
or reference, spectrum which has been corrected for the bandpass
shape, $T_{bas}$ is the mean of the reference signal over the
spectrum, and $T_{sys}$ is the average system temperature as computed
on the standard calibration regions.  The calibrated and LSR velocity
corrected data were imported into AIPS using a modified version of
{\sc otfuv}, {\sc mbfuv}, and gridded using of the task {\sc sdgrd}.
We used a exponential sampling function with a base of 14\min, and a
HPHW of 5\min.  For this aspect of the project the continuum level was
subtracted from the data using off-line channels on both sides of the
line.  All data were then exported to the MIRIAD package and regridded
onto a common grid of Galactic coordinates. MIRIAD and the KARMA
package (Gooch 1995) were used for analysis and visualization.

%----------------------------------------------------------
\section{Results}
\label{sec:res}
%----------------------------------------------------------
We here report on two large \HI\ voids apparent in our data, in the outer
Galaxy near longitude $l=279\deg$.  These voids appear to lie along the edge
of the Carina arm at a galactocentric radius of $\sim 10$ kpc.  Figure
\ref{fig:lv.ps} is a longitude-velocity ({\em l-v}) cut at Galactic latitude
$b=0\deg$ which shows the two voids.  Upon careful examination of the
Kerr (1981, 1986) {\em l-v} diagrams it is clear that the two voids are also
apparent in the latitude averaged sample, and were identified as low density
regions by Kerr (1969).  Below we explore the physical properties of the
voids, determine that they are \HI\ shells, and hypothesize as to their
origins.

%-------------------------
\subsection{Shell 1: GSH 277+0+36  --- Morphology}
\label{subsec:sh1-morph}
%-------------------------
The first void is at $v=+36~{\rm km~s^{-1}}$, $l=277$\deg, $b=0$\deg.  This
void is extremely dramatic with brightness temperatures on the order of 3 K
in the center, whereas the brightness temperatures at the edges are on the
order of 50 K.  In addition, the void is apparent over a large range of
velocities from $v\approx +15$ \kms\ to $v\approx +55$ \kms.  Figure
\ref{fig:chan.ps} shows a grey-scale representation of the channel maps over
the velocity range of both holes.  Every second velocity channel is
displayed for $+9.5$ \kms\ $\leq v \leq +87$ \kms.  There are several
noteworthy features related to this void.  The first is the development of
the shell around $v=+13$ \kms\ at $l=276.5\deg$ and $b=0\deg$.  This feature
starts as a small ring of emission and quickly grows in successive velocity
channels to form the large void with brightened edges in the center of the
maps.  Second, there is another shell-like structure at lower latitudes,
centered on $l\approx 278.2\deg$ and $b\approx -3\deg$ from $v=+23$ \kms\ to
$v=+33$ \kms.  This shell appears to join with the larger, more prominent
void by $v=+26$ \kms, forming one large shell.  Third, the shell is not
entirely closed at the top and bottom.  There are several filamentary
extensions which extend both above and below the shell.  Fourth, around
$v=+49$ \kms\ the back cap of the largest shell becomes visible as the strong
emission in the channel maps.  The full shell has an angular diameter of
$\sim 5\deg.7$ in longitude and $\sim 3\deg.5$ in latitude, with
extensions to $|b|> 10\deg$.

Using a standard rotation model for the Galaxy (Fich, Blitz and Stark 1989)
we determine a kinematic distance to the shell based on the center velocity
of $D=6.5\pm 0.9~{\rm kpc}$ (see Figure~\ref{fig:profile1.ps}), and a
galactocentric radius of $R_g \approx 10~{\rm kpc}$.  Using this distance we
determine that the shell has a radius $R_{sh} = 305\pm 45~{\rm pc}$,
classifying it among the largest shells, or ``supershells'' in our Galaxy
(Heiles 1984).

Perhaps the most interesting morphological features of the shell are the
apparent break-outs in Galactic latitude which extend beyond our latitude
coverage to at least $b=\pm 10\deg$, or $\sim 1.1$ kpc at the distance of
the shell. Figures \ref{fig:chan.ps} and \ref{fig:slice1.ps} clearly show
several channels to upper latitudes.  Figure \ref{fig:slice1.ps} is a
composite of three orthogonal slices through the data cube at the position
marked by the cross.  At the bottom is the {\em l-v} cut, the
latitude-longitude ({\em l-b}) cut with the marked position is in the center,
and to the right is the latitude-velocity ({\em b-v}) slice.  There are at
least two northern channels to the halo visible in the plane of the sky, at
least two along the line of sight, and at least one complete southern
channel, plus a possible second southern channel.  The second southern
channel may be capped as part of the low latitude shell structure noted
above.  It also appears that the shell is slightly inclined with respect to
the line of sight.  The southern chimneys are particularly visible in the
early channels (19 $\leq v \leq 35$ \kms), whereas the northern chimneys are
more dominant in the later channels.  This effect is also visible in the
{\em b-v} slice in Figure \ref{fig:slice1.ps}.  The extended latitude
morphology strongly suggests that the supershell has in fact exceeded the
scale height of the Galaxy and is producing a ``chimney'' into the halo.

The shell's several small channels to the upper layers are much more
reminiscent of Galactic ``worms'' (KHR 1992), than of the large ``cone'', or
``mushroom'' shapes detected by Normandeau \etal (1996) and Mashchencko \etal
(1999), respectively.  In fact, KHR catalogued two Galactic worm candidates
which may be associated with the chimney edges.  GW 274.7+2.7 at $31.2$
\kms\ $\leq v \leq 45.4$ \kms, is coincident with the northwestern chimney
edge as marked in Figure~\ref{fig:worms}.  GW 281.5+1.5 is given with
uncertain velocities $40.5$ \kms\ $\leq v \leq 51.5$ \kms, but is coincident
with a feature associated with the northeastern chimney edge at $21$ \kms\ 
$\leq v \leq 29$ \kms.  Because of the large difference in velocities it is
unclear whether these structures are the same.  We do not see, however, any
strong \HI\ features at the position of GW 281.5+1.5 in the velocity range
given by KHR.  Additionally, KHR noted that the \HII\ regions RCW 45 and RCW
46 lie at the base of GW 281.5+1.5, placing them in the eastern edge of the
shell, and RCW 42 in the western edge.

%-------------------------
\subsection{Shell 1: GSH 277+0+36  --- Physical Properties}
\label{subsec:sh1-props}
%-------------------------
In order to better understand the nature of this object we need to know some
of its physical properties such as mass, expansion velocity, and energy
requirements.  There are several ways to estimate the amount of mass swept
up by an expanding shell.  In this discussion, we will explore two of those
methods and compare them with the empirical result of Heiles (1984).  The
first method is to calculate the column density through the center of the
shell, covering the velocities which include the void as well as the shell's
front and back caps.  Using this method we calculated the column density
over the range $12.75$ \kms\ $\leq v \leq 59.74$ \kms\ and used the average
in the center as a representative number for the shell.  We determine an
average column density through the shell center of $N_H = 1.3 \pm 0.1 \times
10^{21}~{\rm cm^{-2}}$.  This value is comparable to other Galactic and
extragalactic supershells (Heiles 1998).  If we assume that the radius of
the shell along the line-of-sight is approximately equal to the radius in
the plane of the sky, then we find that the density of the ambient medium
prior to formation must have been $n(HI)_o \sim 1.2 ~{\rm cm^{-3}}$.  This
value is slightly high for the outer Galaxy, but not unreasonable.  Using
these values, and a factor of 1.4 to account for helium, we determine that
the swept up mass of the shell is $M_{swept} \sim 5.6 \times 10^6~{\rm
M_{\odot}}$.

An alternative way to estimate the swept-up mass is to use column densities
calculated along the shell edges, rather than through the center of the
shell.  In this case, one determines the column density through the edges of
the shell over the range of velocities where the edges are brightened, and
subtracts a baseline column density which is assumed to be representative of
the area into which the shell expanded.  We calculated the average column
density excess along the eastern and western sides of the shell.  For the
baseline number we used a mid-plane position far enough away from the shell
to be independent of the shell walls.  We determined that the mass of these
to be $\sim 1.3 \times 10^6~{\rm M_{\odot}}$.  Assuming that these comprise
roughly half of the total shell mass, we find that the swept up mass of the
shell must be $M_{swept} \sim 2.7 \times 10^6~{\rm M_{\odot}}$, which is
about a factor of two smaller than the previous estimate.

Both mass estimate methods contain possible sources of error.  In the first
method we assume that the Galactic gas at the velocities including the shell
contributes very little to the overall column density through the shell.
Because the brightness temperature in the void is of the order $\sim 3$
K, this is a reasonable assumption, however we are still likely to over
estimate the mass.  The largest source of error with the edge method is in
calculating the baseline.  On the size scale of a large supershell, density
fluctuations attributed to large scale Galactic structure can dramatically
influence baseline estimates.  Also, exactly determining the area of the
shell edge is very subjective, and in this case we may have underestimated
the mass.  On the whole, this method is less reliable than mass estimates
from the center column density.  It is likely, then, that our mass lies
somewhere between the two estimates.  Neither estimate takes the mass of the
chimney walls into account.

Comparing these mass estimates with the masses of shells listed in Heiles
(1979) catalog of \HI\ shells, we find that this supershell is in the top
25\% of the most massive shells in the Galaxy. Using Heiles' (1979) empirical
equation for shell masses based on the radius of the shell: $M \approx 8.5
~R_{sh}^2$ , we find a mass of $\sim 8.7 \times 10^5 ~{\rm M_{\odot}}$ for a
shell of this size, which severely underestimates the column density masses
by as much as 70-85\%.  Since our masses were determined in a similar manner
to Heiles (1979), the severe departure from the global shell characteristics
is surprising.  The shell appears to be extremely massive for its size.

Another important shell characteristic is the expansion velocity.  This
value can help determine the required creation energy and the age.  We
estimated the expansion velocity of this shell in two ways.  First, we made
a velocity profile through the center of the shell, which is compared with
the Galactic rotation curve along that line-of-sight in Figure
\ref{fig:profile1.ps}.  The large trough between $v=14$ \kms\ and $v=58$
\kms\ is the shell.  Using the peaks at either velocity extreme of the shell
we estimate a full velocity width of $\sim 45$ \kms.  Unfortunately it is
difficult to separate the spatial information from the true velocity
information.  However, the velocity gradient for this line of sight is $\sim
10~{\rm km~s^{-1}~kpc^{-1}}$, which corresponds to a velocity spread due to
the spatial extent of only $\Delta v \sim 6$ \kms\ for a spherical shell.
We therefore assume that to first order the expansion velocity is half of
the full velocity width, giving $v_{exp} \sim 22$ \kms.  We also made use of
the tool {\em kshell} in the KARMA visualization package to estimate the expansion
velocity.  {\em Kshell} computes an average brightness temperature on annuli about
a user defined center.  A shell will appear as a half ellipse in the
resultant radius-velocity ({\em r-v}) diagram.  Figure \ref{fig:rv1.ps} is
the {\em r-v} diagram for GSH 277+0+36.  Using the {\em r-v} diagram we can
validate our center, since incorrectly choosing the center will result in a
double ellipse.  We then determine $v_{exp}\sim 20$ \kms, half of the
ellipse width.  The close agreement between the {\rm r-v} diagram expansion
velocity, which is essentially an average of all velocity profiles through
the shell, and the single velocity profile through the shell center is
comforting.  Finally, we must consider the possibility that the shell is not
expanding, but simply a quiescent hole or empty region in the \HI.  In this
case the velocity spread through the center would translate to a
line-of-sight depth of 4 kpc, while the diameter in the plane of the sky is
only 610 pc.

Using the expansion velocity and shell size, we can calculate the expansion
energy the shell, $E_E$.  The expansion energy is defined as the required
amount of energy instantaneously deposited at the center of the shell to
account for the shell's present size and rate of expansion.  Based on
Chevalier's (1974) calculations for supernova remnant expansion, Heiles
(1979) gives a formula for the expansion energy of a shell of radius,
$R_{sh}$ expanding with a velocity of $v_{exp}$ into a medium with ambient
density, $n_o$:
\begin{equation}
E_{E} = 5.3 \times 10^{43}\,\, n_o^{1.12}\,\, R_{sh}^{3.12}\,\, v_{exp}^{1.4}~\,\,\,\, {\rm ergs},
\label{eq:energy}
\end{equation}
where $n_o$ is in ${\rm cm^{-3}}$, $R_{sh}$ is in pc, and $v_{exp}$ is in
${\rm km~s^{-1}}$. Using $n_o=1.2~{\rm cm^{-3}}$, as calculated above, we
find that the expansion energy is $E_{E} \approx 2.4 \times 10^{53}$ ergs.

There is no evidence in the channel maps (Figure~\ref{fig:chan.ps}) that the
shell has been significantly sheared by the effects of differential
rotation.  We can use this fact, the size of the shell, and its expansion
velocity to place limits of the age of the supershell.  From the fact that
the shell does not show dramatic deformation, we estimate that its age is
less than $\sim 20$ Myr (Tenorio-Tagle \& Bodenheimer 1988).  We also
calculate an upper limit on the shell's age of $\sim 15$ Myr using its
present rate of expansion and size.  Given the uncertainties involved in
determining the shell's expansion velocity and whether or not it may have
stalled, we estimate an age in the range 15 - 25 Myr.  It is noted, however,
that this age is very small for an object this large.

%-------------------------
\subsection{Shell 2: GSH 280+0+59}
\label{subsec:sh2}
%-------------------------
Examining the {\em l-v} diagram in Figure \ref{fig:lv.ps} it is clear that
there are two shells which share a common line-of-sight.  In Figure
\ref{fig:chan.ps} one can see this second shell develop around $v=56$ \kms\
at $l=280\deg$, $b=0\deg$.  This shell is much less pronounced than shell 1,
with only a factor of 4 or 5 difference between the rim and the center of
the void.  The effects of limb brightening are not nearly as noticeable,
except on the back edge as seen in the {\em l-v} diagram.  Following the IAU
naming convention, we name the shell, which is centered on $l=280$\deg,
$b=+0.1$\deg\ and $v=59$ \kms, GSH 280+0+59.  Using the center velocity we
determine a kinematic distance of $\sim 9.4$ kpc, and a galactocentric
radius of $\sim 11.5$ kpc.  Given its angular diameter of 2.6\deg, we
calculate a physical radius, $R_{sh} \approx 215$ pc.
Figure~\ref{fig:slice2.ps} shows three orthogonal slices ({\em l-v, l-b,
b-v}) through the cube.  Though it is not clear in the channel maps, it is
quite apparent in the {\em b-v} slice, that this shell is also breaking out
of the disk.  There is one cone-like chimney to the north, and a jet-like
structure to the south of the shell.  These are both most easily seen in the
{\em b-v} slice.  There is is no apparent cap to the chimney which indicates
that it extends to at least 1.4 kpc, far exceeding the \HI\ scale height.

In order to determine the mass of the shell, we calculated the column
density through the center as described above.  Because the shell edges are
much less noticeable than in shell 1, we were unable to accurately determine
an area over which to calculate the column density.  For the averaged center
column density we find $N_{H} = 3.9 \pm 0.7 \times 10^{20}~{\rm cm^{-2}}$.
Again, assuming that the radius along the line-of-sight is comparable to the
radius in the plane of the sky, we find that the ambient density must have
been $n(HI)_o \sim 0.6~{\rm cm^{-3}}$.  This value agrees quite well with
expected values for the outer Galaxy.  Finally, we calculate a swept-up mass
of $M_{swept} \sim 1.1 \times 10^6 ~{\rm M_{\odot}}$.

GSH 280+0+59 shell also appears to be expanding, though with a smaller
velocity than shell 1.  The {\em r-v} diagram in Figure~\ref{fig:rv2.ps}
clearly shows the characteristic half ellipsoidal void for shell 2.  As with
GSH 277+0+36, the velocity gradient is $\sim 10~{\rm km~s^{-1}~kpc^{-1}}$,
so we assume that the expansion velocity is approximately half of the full
velocity width.  Using Figure~\ref{fig:rv2.ps} we determine an expansion
velocity of $v_{exp}\approx 14$ \kms.  The shell is more clearly seen in the
velocity profile (Figure~\ref{fig:profile2.ps}), which lends the result of
$v_{exp}\approx 17$ \kms.  Using equation~\ref{eq:energy} to estimate the
expansion energy, we find $E_E \approx 2.6 \times 10^{52}$ ergs, placing
this shell in the range of moderate energies, easily achievable from the
combined effects of stellar winds and supernovae.

The two shells appear to be interconnected around $v=45$ \kms.  Though the
morphology supports the case that the two shells are associated, it is
unclear whether the association is physical.  On the morphological side of
the argument, one notices in Figures~\ref{fig:lv.ps} and \ref{fig:slice2.ps}
that the wall which separates the two shells is brightest in the longitude
region where they overlap ($279\deg \leq l \leq 277\deg$), and both shells
appear more compressed in that range of longitudes than they do on the
non-interacting side.  This is consistent with expansion slowed by
interaction of the two high density walls.  Physically speaking, though, it
is improbable that the shells could be associated shells if located at the
exact distances implied by their central velocities, as it would imply
line-of-sight extents on the order of kiloparsecs.  However, the errors for
the shell distances are quite large, so that the shell centers may be
brought closer together, to within only 1.2 kpc.  In this case, a slight
elongation of the shells along the line-of-sight could overlap the shells.
Alternately, if one or both of the shells has a systemic velocity which
deviates slightly from its local standard of rest, the shells could also
overlap.  Both possibilities are conceivable, since the shells are
presumably undergoing exaggerated expansion away from the Carina arm, and
the more massive shell could impart a slight systemic velocity on the other
shell.  Related to the latter argument, it is logical to expect shell 2 to
be older than shell 1, and hence able to receive a small kick from shell 1
as it expanded.  The expansion velocity for shell 2 is smaller, and its
edges are much less brightened, implying an older age.  Finally, because
both shells trace the edge of the Carina arm, it would seem likely that they
are associated.

The physical properties for both shells are given in Table~\ref{tab:props}.
Error estimates for the distances are based on our ability to estimate the
center velocity of the shell and random \HI\ cloud motions.  Since both
shells are against the Carina arm it is likely that the shells are
asymmetric in velocity.  We assume that the center velocities are accurate
to within $9~{\rm km~s^{-1}}$ which accounts for an estimated error of 7
\kms\ in determining the center velocity and $6~{\rm km~s^{-1}}$ for random
cloud motions (Dickey 1997).

%-------------------------
\subsection{Comparison to Other Wavebands}
\label{subsec:otherwaves}
%-------------------------
We have obtained publicly available 2.4 GHz continuum (Duncan \etal 1995),
X-ray (Snowden \etal 1995) and far-infrared data (Wheelock \etal 1994) on
this region for comparison with the \HI\ data.  We also obtained CO data
from Grabelsky \etal (1987).  The primary correlation amongst all bands is a
lack of emission in the region 270\deg\ $\leq l \leq$ 285\deg. This result
is consistent, however, with both the line-of-sight traversing a large
distance in between spiral arms, or the line-of-sight crossing a supershell.
        
Using data on the CO ($J = 1 \rightarrow 0$) transition from Dame et
al. (1987), we have calculated the CO column density map for the range of
velocities of shell 1.  The CO column density contours are overlaid on the
$v=36$ \kms\ \HI\ image in Figure~\ref{fig:co1.ps}.  There are several
patches of CO emission which lie on the edge of the shell, and no detectable
emission in the shell interior.  For a shell created by either stellar
winds, or an HVC impact, it is reasonable to expect molecular clouds and
stellar formation in the compressed gas along the shell shock front.  The
most distinct CO feature is a molecular cloud is at $l=279.9\deg$,
$b=-1.6\deg$, $v=35$ \kms\ as given in Table 2 of Grabelsky \etal (1988).
In addition, the lack of CO emission in the shell interior may be as
significant as the clouds seen on the edge, indicating that the region is
devoid of cold, dense gas.

Figure~\ref{fig:2.4GHz} is a map of shell 1 at $v=40$ \kms\ with 2.4 GHz
continuum contours overlaid.  Clearly, emission on the shell edges dominates
over emission in the center.  On the northeastern rim of the shell lies SNR
G279.0+1.1 (Woermann \& Jonas 1988; Duncan \etal 1995).  Woermann \& Jonas
(1988) noted the coincidental position of the strong HI feature at $v=40$
\kms\ and the brighter limb of the SNR.  Using the $\Sigma -D$ relationship
they determined a distance to the SNR of $\sim 3$ kpc.  They concluded that
the SNR could not be associated with the \HI\ feature, as that would place
the SNR at $\sim 8$ kpc and result in a very large physical SNR diameter
($\sim 220$ pc). Because the $\Sigma -D$ relationship is highly uncertain,
we attempted to find a kinematic distance to the remnant.  We searched the
SNR for associated absorption features and morphological matches between the
\HI\ and continuum.  We were unable, however to find any such features, and
therefore cannot conclusively say whether or not there is an association
between the the SNR and the supershell.

We also explored X-ray maps at 1/4 keV, 3/4 keV, and 1.5 keV from Snowden
\etal\ (1992).  In the case that the shell was young we might expect to see
anti-correlated x-rays from the hot interior gas.  The 1/4 keV map
unfortunately has a large instrumental discontinuity through the center of
the shell, making it difficult to determine any characteristics.  However,
1/4 keV X-rays are absorbed by relatively small neutral column densities.
Given that the column density to these shells is $\sim 2 \times 10^{21}~{\rm
cm^{-3}}$, we would not expect to see 1/4 keV X-rays.  X-rays at 3/4 keV
begin to be significantly absorped at column densities of $\sim 10^{21}~{\rm
cm^{-2}}$, therefore we would expect the 3/4 keV map to show significant
attenuation if there were emission from the shell interior.  Neither the 3/4
keV map, nor 1.5 keV map shows any distinct anti-correlation with the \HI\
column density map.

Figure~\ref{fig:100micron} is an {\it IRAS} 5\min\ resolution map of 100
$\mu$m emission in the region with \HI\ column density contours overlaid.
As we would expect, there is good correlation between the \HI\ column
density and the 100 $\mu$m dust emission on the left rim of the shell.  In
addition, there is little emission throughout the shell interior.  The
wispy, fine scale structure in the shell interior traces the outflow
directions well.

%----------------------------------------------------------
\section{Discussion}
\label{sec:disc}
%----------------------------------------------------------
The decrease in \HI\, as well as CO emission, in the region $270\deg \leq l
\leq 280\deg$ and 15 \kms\ $\leq v \leq 50$ \kms\ was previously noted and
identified as an interarm region between the Carina and external spiral arms
(Kerr \etal 1969; Grabelsky \etal 1987).  In the case that this void is an
interarm region its decrease in brightness temperature from arm to interarm
by more than a factor 16 would make it the most pronounced arm edge in the
Galaxy by more than 50\% (Grabelsky \etal 1987).  Previous data, however,
had either poor angular resolution, or averaged over Galactic latitude so
that the shell-like morphology of the void was not apparent.  We now have
the resolution and sensitivity necessary to discern the shell edges. It
appears that the void is not simply an interarm region, but in fact, a
galactic supershell.  The shell edges curve around to partially close the
shell at the top and bottom.  In addition, the shell appears limb brightened
on all edges, suggesting gas compression due to the shell's expansion.  Both
of these traits, as well as the observed expansion, are inconsistent with an
interarm interpretation.  The bowl shape of the velocity profile through the
center of the shell (Figure~\ref{fig:profile1.ps}) is indicative of a shell,
as well.  Finally, there are breakouts, where the shell appears to be
blowing gas up into the Galactic halo.  All of the morphological evidence
strongly supports the conclusion that the gas has been displaced, both
parallel to and out of the plane.

Based on the size, energy requirements, and positions of these shells we
explore possible formation methods.  GSH 277+0+36 is difficult to understand
because of its large energy requirements, relatively large size and mass,
and unusual position adjacent to the Carina tangent.  As suggested by many
authors (Heiles 1984, Rand \& van der Hulst 1993) it is difficult to
envisage a shell with expansion energies in excess of $10^{53}$ ergs created
by the combined effects of SNe and stellar winds.  In this particular case
if we use 15 Myr as an upper limit to the age, assume SNe with energies of
$\sim 10^{51}$ ergs, then we would expect a supernova rate on the order of
one every $6 \times 10^4$ yrs in the progenitor OB association, which is
about four times higher than suggested by Tomisaka \& Ikeuchi (1986).  For
the Galaxy as a whole, the supernova rate is about one every fifty to a
hundred years (Cappellaro \etal 1999), or $\sim 10^{-13}~{\rm
SNe~pc^{-3}~yr^{-1}}$.  GSH 277+0+36 supershell would require a supernova
rate of the same order as the Galactic rate.  It is not reasonable to expect
the supernova rate in an interarm region of the outer Galaxy to be as high
as the rate for the Galaxy as whole (unless there is a star cluster in the
interarm region, which might be possible if a molecular cloud survives the
arm/interarm transition).

A search of OB association catalogs reveals no associations in the
neighborhood of either GSH 277+0+36 or GSH 280+0+59 (Humphreys 1978; Mel'nik
\& Efremor 1995).  However, the majority of OB associations listed in these
catalogs are restricted to a 3~kpc radius from the Sun, so it is unlikely
that an OB association at the distance of GSH 277+0+36 would have been
catalogued.  Finally, as mentioned above, the only known SNR in the region
is G279.0+1.1 (Duncan \etal 1992), for which there is no conclusive
association with the shell.  For a large majority of supershells, however,
no OB associations or supernova remnants (SNRs) have been associated.

The estimate of the initial \HI\ number density in the region of $n_o(HI)
\sim 1.2~{\rm cm^{-3}}$ is also inconsistent with values expected for
interarm regions.  It would seem that there has been mass influx to boost
the number density of the ambient medium.  On the basis of all these
difficulties with formation theory as a result of stellar winds and
supernovae in an interarm region, we are forced to question whether the
supershell may have formed differently.  One can think of two possible
alternative explanations.  First, that the shell was formed as a result of a
high velocity cloud (HVC) impact.  Second, that the shell actually formed in
the edge of the Carina spiral arm then expanded into and widened the
interarm region.

We first consider the possibility that the shell was formed by the impact of
a HVC with the Galactic disk.  This possibility has been suggested by
numerous authors as an alternative way to reach the high energy requirements
of supershells (Tenorio-Tagle 1980; Heiles 1984; Tenorio-Tagle \etal 1987).
Several HVC-Galactic disk collisions have been hypothesized in external
galaxies.  The most thoroughly explored of these is the large supershell in
NGC 4631 (Rand \& van der Hulst 1993; Rand \& Stone 1996), which is believed
to have been formed with an input energy on the order of $10^{55}$ ergs.
The extremely large energy demands of this shell required an alternative
explanation to SNe and stellar winds.  Another argument in favor of
cloud-disk collisions is that they can occur at any place in the Galaxy, and
hence overcome the problem of large stellar population dependent shells.  In
the case of GSH 277+0+36, which is located in a region of low stellar
population density, a cloud-disk collision seems reasonable.  A cloud-disk
collision could also result in a deposition of cloud mass in the region,
resulting in the anomalously high ambient density calculated.  However, one
does not necessarily expect the morphology of an HVC-disk impact to resemble
the morphology of GSH 277+0+36.  Though models show a spherical shock
developing for HVCs travelling at low enough velocities, the nearly closed
edges combined with multiple channel-like extensions is inconsistent with
models (Tenorio-Tagle \etal 1987).

An alternative, and perhaps more plausible, explanation is that the shell
was formed at the edge of the Carina spiral arm, and therefore widened the
interarm region to appear as though the shell actually formed in between
spiral arms.  In this scenario, it is not impossible to imagine stellar
populations dense enough in the spiral arm to provide the $\sim 250$
supernova producing OB stars required to make the shell.  In addition, the
intertwining shells that make up GSH 277+0+36 indicate that there may have
been more than one wave of star formation.  In which case, it is much more
likely that there would have been enough energy to create this shell with
supernovae and stellar winds.  Furthermore, if the shell did form in the
edge of the spiral arm the energy requirements would be decreased.  The
Chevalier (1974) expansion energy equation assumes expansion into a
relatively uniform medium. However, a shell expanding into a lower density
region can attain a larger radius and expansion velocity than one expanding
into a constant, higher density region.  If the ambient medium density
dropped by a factor of $\sim 2$ at the edge of the arm, then shell expansion
would be accentuated in the direction of the interarm region by as much as
15\%.  The shell would also not decelerate as quickly, resulting in a larger
measured expansion velocity.  These two factors lead to a calculated
expansion energy which may be as much as a factor of two higher than the
actual energy required to create this shell.  It is important to note,
however, that the energy would still fall within the calculated errors.

%----------------------------------------------------------
\section{Conclusions}
\label{sec:concl}
%----------------------------------------------------------
In conclusion, we have found two large HI shells in the outer Galaxy.  The
first and most dramatic, GSH 277+0+36, can be classified as a supershell on
the basis of its large size and expansion energy.  Prior interpretation of
this large void as an interarm region now seems inappropriate on the basis
of the supershell's chimney and shell-like morphology.  The supershell most
probably exists in the region between spiral arms, though it was not
necessarily formed there.  The strong arm-interam contrast previously
noticed has undoubtedly been enhanced by the supershell edges.  We find
evidence for molecular clouds along the supershell's edges, indicating that
star formation may have been initiated by the supershell's expansion.
Because of the shell's unusual position between spiral arms, and its large
formation energy requirements we have considered several formation theories
for this shell.  We have considered the conventional formation method of
stellar winds and supernovae, an HVC collision with the Galactic disk, and
finally we have raised the possibility that the shell formed in the Carina
arm and expanded into the interarm region.  We believe that the latter is
the most likely scenario, as it decreases the energy requirements and is
consistent with theories of Galactic structure which predict higher star
formation rates and therefore higher supernova rates in the spiral arms.

The second shell, GSH 280+0+59, though smaller than the first shell, is
large by Galactic standards with $R_{sh} \sim 220$ pc.  It also appears to
have blown out of the Galactic plane.  While there is no definitive
interaction between the two shells, it is possible that they may be
interacting if one has a systemic velocity which departs from its local of
standard of rest by $\sim 20$ \kms.  They appear, however, to be distinct
shells which presumably formed independently.  The energy requirements for
the smaller shell are much more reasonable, indicating that the shell could
have been created by $\sim 20$ supernovae, or equivalent stellar winds, over
several million years.

The effects of these shells on their local ISM is dramatic.  Regardless of
whether they are associated and whether or not GSH 277+0+36 formed in the
Carina arm, they have significantly reshaped the large scale structure of
the Galaxy in that region on the timescale of millions of years.  Because
other galaxies are so dramatically influenced by shells, supershells, and
chimneys, it is reasonable to expect that the Milky Way has been similarly
influenced.  However, the catalogued shells and chimneys have not revealed
the level of influence on the structure of the Milky Way as those seen in
the Large and Small Magellenic Clouds.  In addition, the relatively few
chimneys seen cannot support the halo.  We expect, therefore, that there are
many more supershells and chimneys to be detected as we probe deeper into
the Galaxy with the Southern Galactic Plane Survey.  It is imperative to
understanding the structure of the Galaxy that we have a complete catalog of
supershells and chimneys, particularly in the inner Galaxy.

\acknowledgements The authors would like to thank L.\ Staveley-Smith and J.\
E.\ Reynolds for their assistance with the observations.  We also thank T.\
Dame for providing the CO data cube.  JMD and NMM-G acknowledge support of
NSF grant AST-9732695 to the University of Minnesota.  NMM-G is supported by
NASA Graduate Student Researchers Program (GSRP) Fellowship NGT 5-50250.  BMG
acknowledges the support of NASA through Hubble fellowship grant
HF-01107.01-98A awarded by STScI, which is operated by AURA Inc. for NASA
under contract NAS S-26555.
%--------------------------------------------
%Bibliography
%--------------------------------------------
%---------------------------------------------
%Figures
%---------------------------------------------
\newpage
\vfil\eject
%\centering {\psfig{figure=f1.eps}} 
\figcaption[f1.eps]{A grey-scale {\em l-v} diagram at
$b=0\deg$ showing the junction of two shells.  This is a cut at $b=0$\deg.
The grey scale is linear and runs from 0 K (white) to 35 K (black).
\label{fig:lv.ps}}

%\centering {\psfig{figure=f2a.ps,height=7.5in}} 
\figcaption[f2a.ps]{Velocity channels from $v=10~{\rm
km~s^{-1}}$ to $v=88~{\rm km~s^{-1}}$ in steps of $1.6~{\rm km~s^{-1}}$.
The velocity is given in the upper left-hand corner of each plane.  The grey
scale is linear and runs from 0 (white) to 40 K (black), as shown on the
wedge at the side.
\label{fig:chan.ps}}

%\newpage
%\centering {\psfig{figure=f2b.ps,height=7.5in}} 
\setcounter{figure}{1}
\figcaption[f2b.ps]{Continued
\label{fig:chan2.ps}}

%\newpage
%\centering {\psfig{figure=f2c.ps,height=5.625in}} 
\setcounter{figure}{1}
\figcaption[f2c.ps]{Continued
\label{fig:chan3.ps}}

%\newpage
%\centering {\psfig{figure=f3.ps,height=6in}} 
\setcounter{figure}{2} \figcaption[f3.eps]
{A slice through GSH 277+0+36, at $l=277.97\deg$, $b=-0.309\deg$.  The
shell's center velocity and walls are marked on the profile.  Below the
profile is the rotation curve for that line-of-sight (Fich, Blitz, \& Stark
1989).
\label{fig:profile1.ps}}

%\newpage
%\centering {\psfig{figure=f4.ps,height=6in}} 
\figcaption[f4.eps]{This image is a composite of three
orthogonal slices through the data cube.  The position of the slices is
marked by a plus sign in all three planes at $l=276.1$\deg, $b=0$\deg,
$v=35$ \kms.  The grey scale is linear and runs from 0 (white) to 35 K
(black).
\label{fig:slice1.ps}}

%\newpage
%\centering {\psfig{figure=f5.ps,width=4.5in}} 
\figcaption[f5.ps]{A velocity slice at $v=32.54$ \kms.  The
arrow indicates the position of an \HI\ feature that is coincident with GW
274.7+2.7 from Koo \etal 1992.
\label{fig:worms}}

%\newpage
%\centering {\psfig{figure=f6.ps,width=5in}} 
\figcaption[f6.eps]{This image is a {\em r-v} diagram
created using the {\em kshell} tool of the KARMA package.  The void is visible as
the light ellipse in the center of the image surrounded by the shell
emission.  The grey scale is linear and runs from 0 to 35 K, as shown in the
color bar to the right.
\label{fig:rv1.ps}}

%\newpage
%\vspace{-0.5in}
%\centering {\psfig{figure=f7.ps,height=6in}} 
\figcaption[f7.eps]{This image is a composite of three
orthogonal slices through the data cube.  The position of the slices is
marked by a plus sign in all three planes at $l=280$\deg, $b=0$\deg,
$v=59.74$ \kms.  The grey scale is linear and runs from 0 (white) to 35K
(black).
\label{fig:slice2.ps}}

%\newpage
%\centering {\psfig{figure=f8.ps,width=5in}} 
\figcaption[[f8.eps]{This image is a {\em r-v} diagram
created using the {\em kshell} tool of the KARMA package.  As in
Figure~\ref{fig:rv1.ps}, the void is visible as the light ellipse just above
the center of the image, surrounded by the shell emission.  The grey scale
is linear and runs from 0 to 35 K.
\label{fig:rv2.ps}}

%\newpage
%\centering {\psfig{figure=f9.ps,height=5in}} 
\figcaption[f9.eps]{A slice through GSH 280+0+66.  The
shell's center velocity and walls are marked on the profile.  Below
the profile is the rotation curve for that line-of-sight (Fich, Blitz, \&
Stark 1989).
\label{fig:profile2.ps}}

%\newpage
%\centering {\psfig{figure=f10.eps,width=4in,rotate=270}} 
\figcaption[f10.eps]{A velocity slice at $v=39$ \kms\ with
CO column density contours overlaid.  The column density was calculated for
the range of velocities spanning the middle of GSH 277+0+36 ($33.2$ \kms
 $\leq v \leq 38.4$ \kms).  Contour levels go from 5 K \kms\ to 40 K \kms\ in
intervals of 2 K \kms.
\label{fig:co1.ps}}

%\newpage
%\centering {\psfig{figure=f11.ps,width=4in}} 
\figcaption[f11.ps]{2.4 GHz continuum emission contours
from Duncan \etal (1995) overlaid on the $v=40$ \kms\ \HI\ greyscale image.
The contours start at 260 mK, with intervals of 130 mK.
%The contours start at 160 mJy~${\rm Bm^{-1}}$, with intervals of 80
%mJy~${\rm Bm^{-1}}$.
\label{fig:2.4GHz}}

%\newpage
%\centering {\psfig{figure=f12.ps,height=4.5in}} 
\figcaption[f12.ps]{The grey scale shows {\it IRAS}
100$\mu$m data.  The scale is linear from 0 MJy~${\rm sr^{-1}}$ to
115MJy~${\rm sr^{-1}}$.  Overlaid are \HI\ column density contours from 200
K \kms\ to 2000 K \kms\ with 100 K \kms\ intervals.  The column density was
calculated over the range of velocities representing the interior of GSH
277+0+36 ($24.9$ \kms $\leq v \leq 42.4$ \kms).
\label{fig:100micron}}

%%---------------------------------------
% Table of Shell Properties
%---------------------------------------
\newpage
\begin{table}[htb]
\centering
\begin{tabular}{lcc}
\tableline
\tableline
Shell Property & Value  for Shell 1 & Value for Shell 2\\
\tableline
Center ($l$,$b$) (deg)  . . . . . . . . . . . . . & 277.5, 0.0 & 279.8, 0.1\\
Center velocity (\kms) . . . . . . . . . & 36 & 59\\
Distance (kpc) . . . . . . . . . . . . . . . & $6.5 \pm 0.9$ & $9.4 \pm 0.9$\\
Radius (pc). . . . . . . . . . . . . . . . . & $305 \pm 45$ & $215 \pm 20$\\
Galactocentric radius (kpc) . . . . . . . & $10.0 \pm 0.2$ &
$11.6\pm 0.3$\\
Expansion velocity (\kms). . . . . . . & $20\pm3$ & $15\pm2$\\
Swept-up mass ($10^6 M_{\odot}$)  . . . . . . . . . & 2.7 - 5.6 & $1.1\pm 0.2$ \\
Ambient density (${\rm cm^{-3}}$) . . . . . . . . . & $1.2 \pm 0.1$ & $0.6 \pm 0.1$\\
Expansion energy ($10^{52}$ ergs) . . . . . . & $24\pm12$  & $2.6\pm 1.0$ \\ 
\tableline
\end{tabular}
\caption[Shell Properties]{Shell properties.}
\label{tab:props}
\end{table}

\end{document}